\pdfoutput=1
\documentclass[cernpreprint, texlive=2025, UKenglish, texmf, orcidlogo]{atlasdoc}
\usepackage{atlaspackage}
\usepackage{atlasbiblatex}

\usepackage{atlasphysics}
\graphicspath{{figures/}{logos/}}

\usepackage{ANA-HIGG-2019-23-PAPER-defs}

\AtlasTitle{Search for boosted diphoton resonances in the 10 to 70~GeV mass range using
138~fb$^{-1}$ of 13~TeV $pp$ collisions with the ATLAS detector}

\AtlasAbstract{%
A search for diphoton resonances in the mass range between 10 and 70~GeV with the ATLAS experiment at the Large Hadron Collider (LHC)  is presented.
The analysis is based on $pp$ collision data corresponding to an integrated luminosity of 138~fb$^{-1}$ at a centre-of-mass energy
of 13~TeV recorded from 2015 to 2018.
Previous searches for diphoton resonances at the LHC have explored masses down to 65~GeV, finding no evidence of new particles.
This search exploits the particular kinematics of events with pairs of closely spaced photons reconstructed in the detector, allowing examination of invariant masses down to 10~GeV.
The presented strategy covers a region previously unexplored at hadron colliders because of the experimental challenges of recording low-energy photons and estimating the backgrounds.
No significant excess is observed and the reported limits provide the strongest bound on promptly decaying axion-like particles coupling to gluons and photons for masses between 10 and 70~GeV.
}

\author{The ATLAS Collaboration}

\AtlasRefCode{HIGG-2019-23}
\PreprintIdNumber{CERN-EP-2022-183}
\AtlasJournalRef{JHEP 07 (2023) 155}
\AtlasDOI{10.1007/JHEP07(2023)155}
\hypersetup{pdftitle={ATLAS document},pdfauthor={The ATLAS Collaboration}}

\begin{document}

\maketitle

\newpage

\section{Introduction}

One of the primary goals of the Large Hadron Collider (LHC) experiments is to investigate any `beyond the Standard Model' (BSM) scenarios where deviations from Standard Model (SM) behaviour appear in the high-energy regime of observables accessible in hadron collisions.
This work focuses on diphoton final states, looking for new phenomena in the form of narrow resonances $X$ in the mass range between 10 and 70~\GeV, below that targeted in previous diphoton searches at the LHC~\cite{HIGG-2014-04,HIGG-2018-27,CMS-HIG-14-006,CMS-EXO-17-017}, and extending the experimental coverage of the LHC to BSM scenarios where i) all the heavy states are beyond the reach of the LHC, ii) a scalar which is a singlet of the SM gauge group is naturally lighter than the EW scale, iii) the singlet scalar is abundantly produced in proton--proton ($pp$) collisions and iv) the singlet scalar decays promptly into a pair of SM particles, generically with a narrow width.

The natural targets of this search are pseudo Nambu--Goldstone bosons (pNGBs) associated with spontaneously broken approximate global symmetries at the \TeV\ scale or above, often referred to as axion-like particles (ALPs).
A light ALP coupled to gluons would be abundantly produced in $pp$ collisions, and even though it would decay mostly into dijets, its suppressed coupling to photons presents experimental advantages that make the diphoton final state more desirable to look for at the LHC.
Light ALPs with these phenomenological features are predicted by a wide variety of BSM scenarios, such as `heavy' QCD axion models~\cite{Holdom:1982ex,Choi:1988sy,Holdom:1985vx,Dine:1986bg,Flynn:1987rs,Rubakov:1997vp,Choi:1998ep,Berezhiani:2000gh,Hook:2014cda,Fukuda:2015ana,Dimopoulos:2016lvn,Agrawal:2017ksf,Gaillard:2018xgk}, the R-axion in SUSY-breaking models~\cite{Nelson:1993nf,Goh:2008xz,Bellazzini:2017neg} and Composite Higgs models~\cite{Ferretti:2013kya,Belyaev:2015hgo,Ferretti:2016upr}. Light ALPs can also be the mediator of dark-matter freeze out~\cite{Mardon:2009gw} and trigger baryogenesis~\cite{Jeong:2018ucz}. In all of these models, the mass of the ALP is a free parameter without any particularly strong theory expectation, and for this reason the experimental coverage should span all possible masses.

This search uses the diphoton invariant mass distribution as the discriminating variable.
It therefore closely follows the strategy used in previous diphoton searches~\cite{ATLAS-CONF-2018-025,HIGG-2018-27}, but it also exhibits some clear differences.
Selecting photons with transverse energies close to the trigger thresholds sculpts the diphoton invariant mass distribution at its lower edge, limiting the lowest accessible mass and requiring an accurate description of the background shape to avoid drastically limiting the final sensitivity.
This challenge is overcome here by selecting events with pairs of closely spaced photons and large diphoton transverse momentum, typically arising from recoil against a jet.
Concretely, requiring photon pairs with transverse momentum \ptgg\ larger than 50~\GeV\ flattens the background shape, making it easier to describe with simple analytic functions.

A model composed of analytic functions is used to describe the signal and background components and fit the diphoton invariant mass spectrum to search for narrow resonances.
The impact of the uncertainty arising from the analytic function chosen to describe the background shape, estimated by fitting the analytic model to a representative sample of simulated background events, is mitigated by applying a smoothing procedure using Gaussian Processes to the simulated background sample.
In the scenario in which no significant signal excess is observed, limits are placed on the production cross-section times branching ratio, \sfidbr, as a function of the resonance mass $m_{X}$.
Additionally, the observed limits are recast in the ALP parameter space in terms of the mass of an ALP, $m_{a}$, and its decay constant, $f_{a}$.
%


%
%
%

\section{ATLAS detector}

\newcommand{\AtlasCoordFootnote}{%
ATLAS uses a right-handed coordinate system with its origin at the nominal interaction point (IP)
in the centre of the detector and the \(z\)-axis along the beam pipe.
The \(x\)-axis points from the IP to the centre of the LHC ring,
and the \(y\)-axis points upwards.
Cylindrical coordinates \((r,\phi)\) are used in the transverse plane,
\(\phi\) being the azimuthal angle around the \(z\)-axis.
The pseudorapidity is defined in terms of the polar angle \(\theta\) as \(\eta = -\ln \tan(\theta/2)\).
Angular distance is measured in units of \(\Delta R \equiv \sqrt{(\Delta\eta)^{2} + (\Delta\phi)^{2}}\).}

The ATLAS experiment~\cite{PERF-2007-01, ATLAS-TDR-2010-19, PIX-2018-001} at the LHC is a multipurpose particle detector
with a forward--backward symmetric cylindrical geometry and a near \(4\pi\) coverage in  solid angle.\footnote{\AtlasCoordFootnote}
It consists of an inner tracking detector (ID) surrounded by a thin superconducting solenoid providing a \SI{2}{\tesla} axial magnetic field, electromagnetic and hadronic calorimeters, and a muon spectrometer (MS).
The ID covers the pseudorapidity range \(|\eta| < 2.5\).
It consists of silicon pixel, silicon microstrip, and transition radiation tracking detectors.
Lead/liquid-argon (LAr) sampling calorimeters provide electromagnetic (EM) energy measurements with high granularity.
A hadron sampling calorimeter composed of steel and scintillator tiles covers the central pseudorapidity range (\(|\eta| < 1.7\)).
The endcap and forward regions are instrumented with LAr calorimeters for EM and hadronic energy measurements up to \(|\eta| = 4.9\).
For \(|\eta| < 2.5\), the EM calorimeter is divided into three longitudinal layers, which are finely segmented in $\eta$ and $\phi$, particularly in the $\eta$ direction in the first layer.
This segmentation allows the measurement of the lateral and longitudinal shower profile, and the calculation of shower shapes~\cite{EGAM-2018-01} used for electron and photon identification.
The longitudinal segmentation of the EM calorimeter is also exploited to calibrate the energy response of photon candidates~\cite{EGAM-2018-01}.
The MS surrounds the calorimeters and is equipped with three large air-core toroidal superconducting magnets with eight coils each.
The field integral of the toroids ranges between \num{2.0} and \SI{6.0}{Tm} across most of the detector.
A two-level trigger system is used to select events to be recorded~\cite{TRIG-2016-01}.
The first-level trigger is implemented in hardware and uses a subset of the detector information.
It selects events of interest within the event rate limitation of \SI{100}{\kHz}.
This is followed by a software-based trigger which runs algorithms similiar to those in the offline reconstruction software, reducing the event rate to approximately \SI{1}{\kHz}.
An extensive software suite~\cite{ATL-SOFT-PUB-2021-001} is used in data simulation, in the reconstruction and analysis of real and simulated data, in detector operations, and in the trigger and data acquisition systems of the experiment.


%
%
%

\section{Data and simulated event samples}
\label{sec:data_mc}

The search is performed using the $\sqrt{s} = 13$~\TeV\ $pp$ collision dataset with a bunch spacing of 25~ns collected from 2015 to 2018 by the ATLAS detector, referred to as the full Run~2 dataset in the following.
Only events with stable beam conditions and all ATLAS subsystems operational are considered~\cite{DAPR-2018-01}, corresponding to an integrated luminosity of 138~\ifb~\cite{ATLAS-CONF-2019-021}.
The data were recorded using a diphoton trigger~\cite{TRIG-2018-05} that required two electromagnetic energy clusters satisfying identification criteria based on the expected shape of the two electromagnetic showers and with transverse energies \ET\ above a certain threshold that varied across the data-taking period to cope with the increase in instantaneous luminosity over the years.
In 2015 and the first portion of 2016, the \ET~threshold was 20~\GeV, while in the remainder of 2016 an \ET~threshold of 22~\GeV\ was imposed.
During 2017 and 2018, the \ET~threshold was lowered to 20~\GeV\ and combined with an additional requirement on the calorimetric isolation transverse energy.
Data collected with alternative and prescaled diphoton triggers with looser identification criteria for the EM shower shapes, corresponding to an integrated luminosity of $6.3$~\ifb,  are used for background estimation purposes.

Monte Carlo (MC) simulated events are used to optimize the analysis selections and to characterize the signal and background shapes.

Signal event samples were generated for a hypothetical resonance produced in gluon--gluon fusion in association with up to two additional jets for resonance masses between 10 and 80~\GeV.
The decay width $\Gamma_{a}$ was set to 4~\MeV, negligible compared to the experimental resolution, to describe a hypothetical resonance in the narrow-width approximation (NWA).
The samples were generated using the effective-field-theory approach~\cite{Artoisenet:2013puc} implemented in \MGNLO~\cite{Alwall:2014hca} with the \NNPDF[2.3lo] PDF set~\cite{Ball:2014uwa}, and using the A14 set~\cite{ATL-PHYS-PUB-2014-021} of tuned parameters and  \PYTHIA[8.240]~\cite{Sjostrand:2007gs} to simulate parton showering, hadronization and the decay of the resonance into a pair of photons.

Background events with two prompt photons and associated jets were simulated using the \SHERPA[2.2.4]~\cite{Gleisberg:2008ta,Bothmann:2019yzt} event generator.
Matrix elements were calculated in perturbative QCD (pQCD) at next-to-leading order (NLO) for up to one additional parton, and at LO for two or three partons, and merged with the \SHERPA parton-shower simulation using the \MEPSatNLO prescription~\cite{Hoeche:2011fd,Hoeche:2012yf,Catani:2001cc,Hoeche:2009rj}.
The \NNPDF[3.0nnlo] PDF set was used in conjunction with a dedicated parton-shower tune in the \SHERPA generator.
Interference effects between the resonant signal and all background processes are expected to be small for narrow-width signals and are neglected in this analysis.

The effects of multiple $pp$ interactions in the same bunch crossing as the hard scatter and in neighbouring ones (defined as pile-up) are included using simulated events generated with \PYTHIA[8].
Simulated events were weighted to reproduce the distribution of the average number of interactions per bunch crossing observed in data.

All simulated signal events were processed using a full simulation of the ATLAS detector~\cite{SOFT-2010-01} based on \GEANT~\cite{Agostinelli:2002hh}.
The background \gamgam~events were processed using a fast simulation of the ATLAS detector~\cite{ATL-PHYS-PUB-2010-013}, where the full simulation of the calorimeter is replaced with a parameterization of the calorimeter response.
All simulated events were reconstructed with the same reconstruction algorithms as those used for data.


%
%
%

\section{Object and event selection}
\label{sec:selection}

Photon candidates are reconstructed from topological clusters of energy deposited in the EM calorimeter, as well as from charged-particle tracks and conversion vertices reconstructed in the inner detector, and they are calibrated as described in Ref.~\cite{EGAM-2018-01}.
The event selection requires at least two photon candidates with transverse energies larger than 22~\GeV\ and $|\eta|< 2.37$, excluding the barrel-to-endcap transition regions of the calorimeter, $1.37 <|\eta|< 1.52$.
The transverse energy requirement is chosen to mitigate the effect of the trigger efficiency turn-on from the trigger thresholds discussed in Section~\ref{sec:data_mc}.
The properties of the EM clusters associated with the two highest-\ET~photons and additional information from the tracking systems are used to identify the diphoton production vertex~\cite{HIGG-2013-08}, which is used to correct the photon direction, resulting in improved \mgg~resolution.

To reduce the background from jets, photon candidates are required to satisfy \emph{tight} identification criteria based on the shape of EM showers in the LAr calorimeter and energy leakage into the hadronic calorimeter~\cite{EGAM-2018-01}.
Events with one or both photon candidates passing a looser identification are kept for background estimations.
The \emph{tight} identification is optimized in ranges of photon \ET\ and $|\eta|$, and has an identification efficiency that increases with \ET from 70\% at \SI{22}{\GeV} to 90\% above \SI{50}{\GeV}.

To further improve the rejection of jets misidentified as photons, the candidates are required to be isolated using information from both the calorimeter and tracking subsystems.
The calorimeter isolation transverse energy \ETcalo~is required to be smaller than $0.065\ET^\gamma$, where \ETcalo~is defined as the sum of the transverse energies of positive-energy topological clusters~\cite{PERF-2014-07} within a cone of size $\Delta R = 0.2$ around the photon candidate, excluding the photon transverse energy $\ET^\gamma$ and correcting for pile-up and underlying-event contributions~\cite{STDM-2010-08,Cacciari:2008gn,Cacciari:2009dp}.
The track isolation transverse energy \ETtrk~is required to be less than $0.05\ET^\gamma$, where \ETtrk~is defined as the scalar sum of the transverse momenta of tracks with $\pT > 1$~\GeV\ in a $\Delta R=0.2$ cone around the photon candidate, and which satisfy some loose track-quality criteria, are not associated with a photon conversion, and originate from the diphoton production vertex.
The combined isolation efficiency for pairs of photons fulfilling the identification requirement in simulated signal samples increases with \mgg~from 80\% at \SI{10}{\GeV} to 90\% at \SI{90}{\GeV}.

The diphoton invariant mass is computed using the transverse energies of the leading and subleading photon candidates and their angular separation in both azimuth $\phi$ and pseudorapidity $\eta$, determined from their positions in the calorimeter and the diphoton production vertex.

An additional kinematic selection is placed on the transverse momentum of the diphoton system, \ptgg, requiring events to have a diphoton pair with \ptgg~> 50~\GeV.
This requirement is motivated by the fact that the analysis targets diphoton pairs with low masses, down to about half the trigger energy thresholds, and such pairs are typically highly boosted with respect to the ATLAS detector rest frame.
The \ptgg~requirement is chosen in order to reach the best compromise between the statistical uncertainty in the lowest part of the spectrum and sculpting effects on the background shape from the trigger efficiency turn-on, the modelling of which would result in large systematic uncertainties.

In total, 1\,166\,636 data events with $\mgg<80$~\GeV\ are selected.

Following the detector-level selection, the measurement of the signal production cross-section is performed in a fiducial volume defined from the simulated samples by requiring two photons at particle level with $\ET>22$~\GeV, $|\eta|<2.37$ and $\ptgg>50$~\GeV.
The particle isolation, defined as the scalar sum of the \pT\ of all the stable particles (except muons and neutrinos) found within a $\Delta R=0.2$ cone around the photon direction, is required to be less than $0.05\ET^\gamma$.
This isolation requirement is chosen to reproduce the detector-level selection.

%
%


%
%
%

\section{Signal modelling}
\label{sec:signalModelling}

The shape of a possible signal in the diphoton invariant mass distribution is modelled by a double-sided Crystal Ball (DSCB) function, composed of a Gaussian core with power-law tails~\cite{HIGG-2014-04,Oreglia:1980cs}, whose parameter values evolve linearly with respect to the mass \mX.
The parameters of the DSCB function are extracted from fits to the \MADGRAPH~simulated signal samples.
The width of the Gaussian core is entirely determined by the resolution of the detector and ranges from 0.2 to 1.2~\GeV, as shown in Figure~\ref{fig:sigma_param}.
Good agreement between the signal parameterization and the simulated signal samples is found, with differences below 1\% of the fitted signal yield.
An example of the simulated resonance overlaid with the signal model parameterization is shown in Figure~\ref{fig:signal_point30}.

\begin{figure}[htb]
\centering
\subfloat[]{
\includegraphics[width=0.47\textwidth]{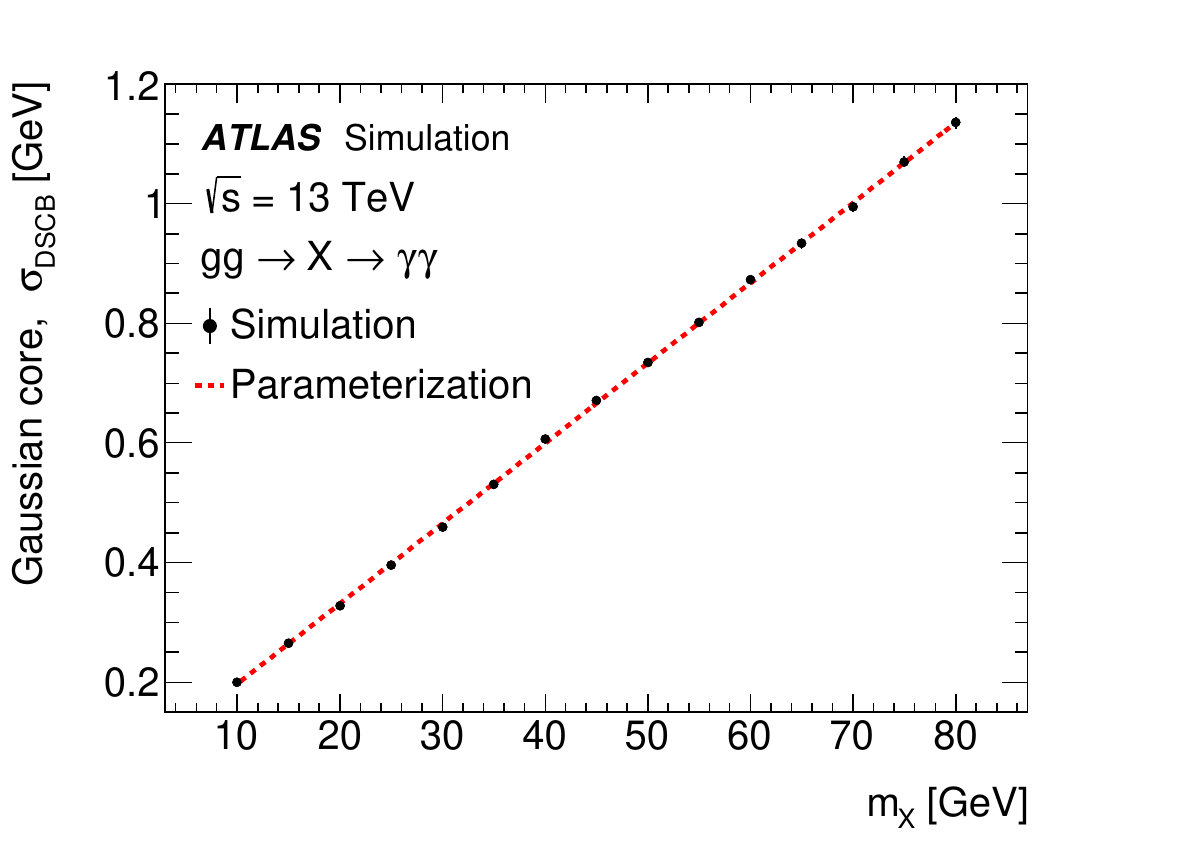}\label{fig:sigma_param}}
\hfill
\subfloat[]{
\includegraphics[width=0.47\textwidth]{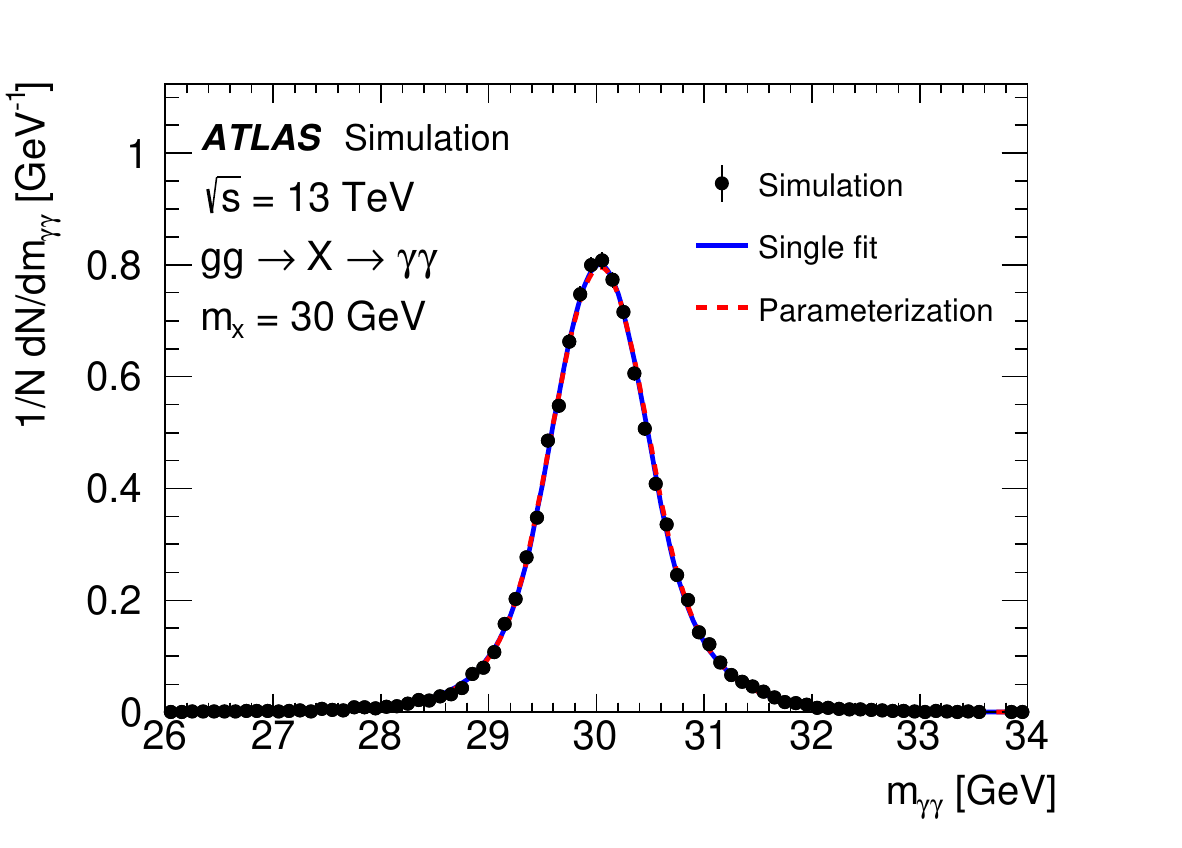}\label{fig:signal_point30}}
\hfill
\caption{
(a) Width of the Gaussian core of the DSCB  function as a function of \mX\ (solid markers) overlaid with the linear parameterization (dashed line). The detector resolution is 2.0\% at 10 \GeV~and 1.4\% at 80 \GeV.
(b) Simulated diphoton invariant mass distribution of a narrow-width signal resonance with $\mX = 30$~\GeV\ (solid markers) overlaid with the DSCB function obtained from a single fit (solid line) and from the signal model parameterization (dashed line).
}
\end{figure}


%
%
%

\section{Background estimates}
\label{sec:bkgModelling}

The dominant background components consist of continuum \gamgam~production, and of photon--jet ($\gamma j$ and $j \gamma$)\footnote{The order denotes which of the two, either the photon or the jet, has a larger transverse momentum. In the following, the term $\gamma j$ refers to both contributions.} and jet pair ($jj$) events where one or more jets are misidentified as photons.
Other backgrounds arising from electrons faking photons in $Z$ boson decays are found to be negligible in the mass range of this search and are not considered.
The analysis makes use of a data-driven background estimate in which the continuum background shape is parameterized by an analytic function.
The chosen analytic functional form is described in Section~\ref{sec:sub_bkg_model}.
The uncertainty arising from the choice of background model is based on signal-plus-background fits to background-only template histograms with a binning of 100~\MeV, following the methodology described in Ref.~\cite{ATL-PHYS-PUB-2020-028} and further described in Section~\ref{sec:sub_bkg_model}.
The background modelling uncertainty is found to be dominated by the limited size of available simulated event samples.
In order to reduce the impact of the background modelling uncertainty on the analysis, the background templates are smoothed using a Gaussian Processes fit.
This technique is described further in Section~\ref{sec:gp_smoothing}, and its impact on the analysis is detailed in Section~\ref{sec:sub_bkg_uncert}.

\subsection{Background template modelling}
\label{sec:sub_bkg_model}

The background-only template has two components.
The \gamgam~component is built from the simulated \gamgam~samples described in Section~\ref{sec:data_mc}, and the $\gamma j$ and $jj$ components are built from  control samples obtained from data, in which one or both photons must fail the \emph{tight} identification requirements while passing a looser set of identification cuts.
The two components are combined according to their relative fractions.
The relative contribution of each of these processes is shown in Figure~\ref{fig:bkgDecomp} and is  estimated using the two-dimensional sideband method described in Ref.~\cite{STDM-2012-05}.
The purity of the diphoton sample, defined as the fraction of \gamgam~events, increases with \mgg~from 50\% at 10~\GeV\ to 70\% at 80~\GeV\ with an overall uncertainty of 3\% dominated by its statistical component arising from the limited size of the data sample collected with the prescaled triggers described in Section~\ref{sec:selection}.
No significant difference in the diphoton purity is observed between the various LHC data-taking periods of Run~2.

\begin{figure}[htb]
\centering
\includegraphics[width=0.7\textwidth]{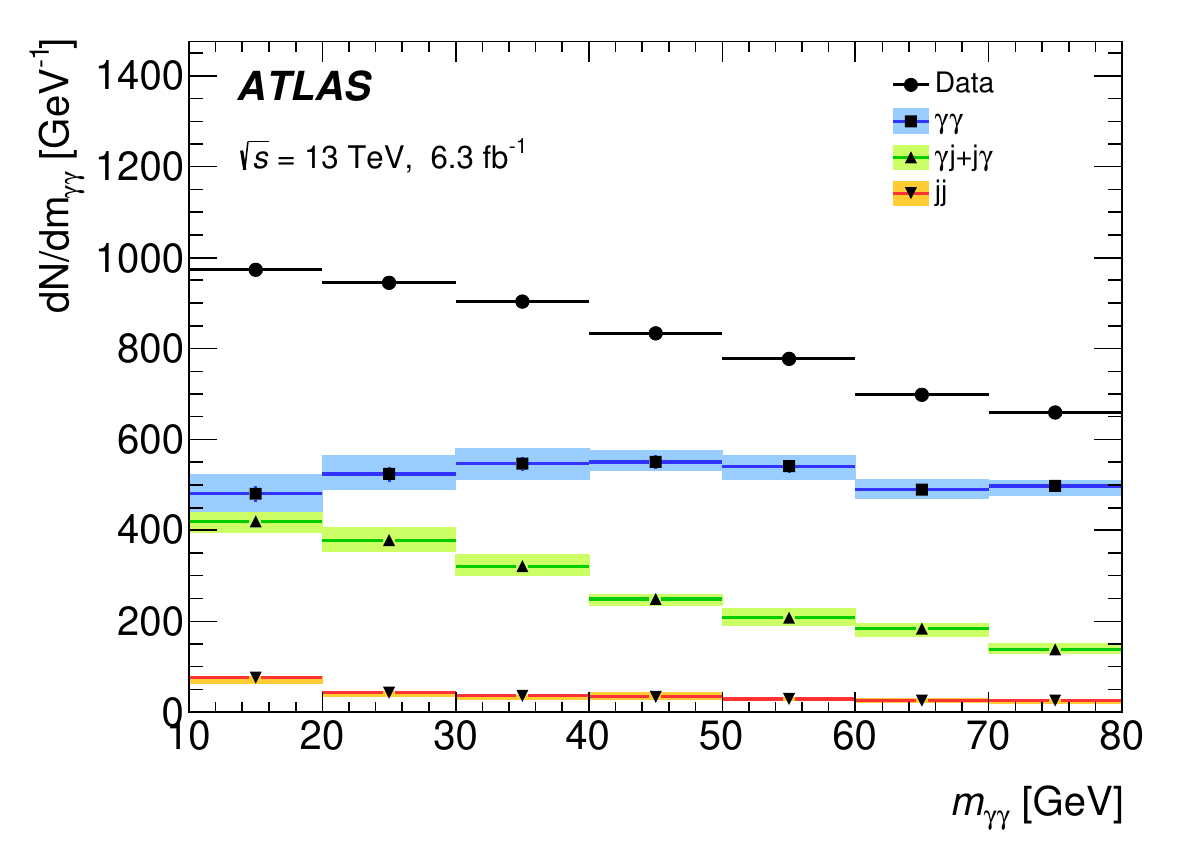}
\hfill
\caption{
Diphoton invariant mass distributions of the data after event selection and their decomposition into contributions from diphoton (\gamgam), photon+jet ($\gamma j$ and $j\gamma $) and dijet ($jj$) events as determined using the two-dimensional sideband method.
The total uncertainties, including statistical and systematic components added in quadrature, are shown as error bands.
}\label{fig:bkgDecomp}
\end{figure}

The goal of this analysis is to reach the lowest invariant mass possible, including the `turn-on' region for masses below 20~\GeV. %
The resulting background shape needs to be described by a more complex analytic form than in previous diphoton resonance searches~\cite{ATLAS-CONF-2018-025,HIGG-2018-27} and is constructed as the combination of two pieces: one capable of describing the turn-on shape and a second used to describe the smoothly decreasing part.
The two-parts analytic function described below was found to adequately model the background shape across the full invariant mass range of the search.

The turn-on region (TO) is described by the following function:
\begin{equation*}
h_\text{TO}(\mgg;f_{0},\tau_\text{TO}) = 1-(1-f_{0})\,\mathrm{e}^{-\frac{\mgg}{\tau_\text{TO}}}\ ,
\end{equation*}
where $f_{0}$ corresponds to the value of the function at $\mgg=0$  and $\tau_{\text{TO}}$ is the length scale of the turn-on.

The smoothly falling region beyond 30 \GeV\ is described by a power-law function multiplied by an `activation' function to increase its flexibility in the high mass region (above 50~\GeV).
For this activation term,  an exponential function times a `Fermi--Dirac-like' function is chosen, and the total function is:

\begin{equation*}
h_\text{High}(m_{\gamgam};c_{1},a_{0},c_{0},\delta_\text{tail},\tau_\text{tail},\delta_\text{thresh},\tau_\text{thresh}) = \underbrace{\left(1-\left(\frac{m_{\gamma\gamma}}{c_{1}}\right)^{a_{0}}\right)^{c_{0}}}_\text{Power-law}\underbrace{\left(1+\frac{\mathrm{e}^{\frac{m_{\gamma\gamma}-\delta_\text{tail}}{\tau_\text{tail}}}}{1+\mathrm{e}^{-\frac{m_{\gamma\gamma}-\delta_\text{thresh}}{\tau_\text{thresh}}}}\right)}_\text{Activation function}\ ,
\end{equation*}
where $c_{1},a_{0}$ and $c_{0}$ are the parameters of the power-law term, and $\delta_\text{tail},\tau_\text{tail},\delta_\text{thresh}$ and $\tau_\text{thresh}$ describe the activation function.
The power-law part is qualitatively described by its endpoints, being 1 at $\mgg=0$, and  0 at $\mgg=c_{1}$, and a fixed value of $c_1=115$~\GeV\ can be set with no impact on the flexibility of the complete model.
The activation function only plays a role above $\delta_\text{thresh}$, below which its value is practically 1.

The complete functional form is obtained by adding the two components and has ten parameters in total:

\begin{equation}
f(m_{\gamgam};\boldsymbol{\theta}_\text{TO},\boldsymbol{\theta}_\text{High},f_{\text{TO}}) = h_\text{TO}(\boldsymbol{\theta}_\text{TO})\cdot f_{\text{TO}} + (1-f_{\text{TO}})\cdot h_\text{High}(\boldsymbol{\theta}_\text{High})\ ,
\label{eq:fullBkg}
\end{equation}
where $f_{\text{TO}}$ is the parameter that describes the relative contribution of the turn-on component and $\boldsymbol{\theta}_{i}$
are the sets of parameters belonging to $h_\text{High}$ and $h_\text{TO}$.
To reduce the potentially large correlations between the ten parameters, a subset of them are fixed to ensure the convergence of the fit.
The choice of free parameters is based on the results of stability tests using generated pseudo-datasets from the best fit to the background template.
The chosen configuration has the largest number of floating parameters, with only three fixed parameters ($c_{1},\delta_\text{thresh},\tau_\text{thresh}$) while the remaining seven parameters are free to vary.

Variations of the nominal background template are built to validate the flexibility of the chosen functional form and subsequently to estimate background modelling systematic uncertainties in Section~\ref{sec:sub_bkg_uncert}.
They are constructed by i) modifying the fractions of the background components, referred to as variations of the \gamgam\ fraction ($f_{\gamgam}$); ii) varying the identification criteria used to define the $\gamma j$ and $jj$ control regions, referred to as variations of the control region; and iii) altering the templates by varying the \ptgg\ cut by 10\%, referred to as \ptgg\ variations.
These variations change the steepness of the turn-on by up to 20\%, and the slope in the high mass region by $\pm5$\%, with respect to the nominal background template.

\subsection{Gaussian Processes to mitigate statistical fluctuations in background templates}
\label{sec:gp_smoothing}

The bias arising from the choice of background model is evaluated from signal-plus-background fits to background-only templates: any fitted signal yield $N_\text{SS}$ is referred to as a `spurious signal' and it is considered as a systematic uncertainty in the modelling of the background shape.
The functional form of Eq.~(\ref{eq:fullBkg}) provides an acceptable maximal spurious signal that is below 30\% of the statistical uncertainty in the mass range between 10 and 75~\GeV, and therefore it is chosen as the background model.

The estimation of this systematic uncertainty requires the shape of the background template to be as close as possible to the shape of the data distribution, but with an event count large enough for the statistical fluctuations of the template to be negligible.
When evaluating the uncertainty, if the background model perfectly describes the representative background sample, then the number of signal events fitted by the signal-plus-background model will be zero.
However, the representative background-only sample for this analysis is constructed using a limited number of simulated diphoton events, and the presence of statistical fluctuations in the sample introduces large statistical fluctuations in the number of fitted signal events, regardless of the quality of the background model.
This issue was addressed previously~\cite{ATLAS-CONF-2018-025} by using simulated datasets with much larger event counts than in data, which leads to smaller statistical fluctuations in the background shape but is computationally expensive.
In order to meet the aforementioned requirements, an alternative approach to easing statistical effects on the background modelling uncertainty by using Gaussian Processes is followed instead.

A Gaussian Process (GP) is a flexible Bayesian machine-learning technique which may be used to obtain a non-parametric fit to an input dataset~\cite{rasmussen}.
The analysis uses the scikit-learn GP implementation~\cite{scikit-learn} to fit a GP to the representative background sample histogram; the posterior mean of the GP fit is used as a smoothed background template.
The combined signal and background model fit is then performed on the smoothed template instead of the original representative background sample. The degree of smoothing applied is controlled through the choice of kernel and its hyperparameters.
A radial basis function (RBF) kernel~\cite{rasmussen} with an additional constant noise component is utilized here.
The RBF kernel includes a length-scale hyperparameter that encodes the correlation between the event counts for different bins in invariant mass.
The contents of bins which are less than the length scale apart in invariant mass are highly correlated, while the contents of those which are much further apart than the length scale are essentially uncorrelated.
Physically, the length scale encodes a minimum feature size expected in the background shape, which in this analysis corresponds to the 1--2~\GeV\ width of the trigger efficiency turn-on region and thus is also greater than the 100~\MeV\ bin width of the original background histogram.
The kernel hyperparameter values are determined in the fit to the representative background sample.
Notably, because GPs are non-parametric in nature, the GP smoothing technique is not expected to significantly bias the shape of the resulting smoothed template towards a specific choice of analytic background model.

GPs may introduce mis-modelling at the edges of the diphoton invariant mass distribution, since edge points are only constrained by their correlation with other data points on one side.
To mitigate these edge effects, the GP fit is performed using an extended invariant mass range of 7--80~\GeV, while the combined signal and background model fit is performed using the nominal analysis invariant mass range of 9--77~\GeV.

Figure~\ref{fig:smooth_toy} shows an example of a pseudo-dataset generated from the nominal background modelling function, as well as the GP-smoothed pseudo-dataset.
The smoothed pseudo-dataset is observed to reproduce the nominal background modelling function shape with relatively high accuracy, and without the bin-scale statistical fluctuations of the original pseudo-dataset.
The smoothed pseudo-dataset shows some oscillatory behaviour beyond the turn-on region; these features are an artefact of the steep slope in the turn-on region pulling the fitted length scale to a smaller value than would otherwise be needed to model the remaining mass range.
Similarly, some mis-modelling of the smoothed pseudo-dataset is observed in the turn-on region because of the length scale being pulled to a larger value by the higher mass region. The magnitude of the fluctuations in the smoothed pseudo-dataset is significantly smaller than that of statistical fluctuations in the unsmoothed pseudo-dataset.

\begin{figure}[h]
\centering
\includegraphics[width=0.70\columnwidth]{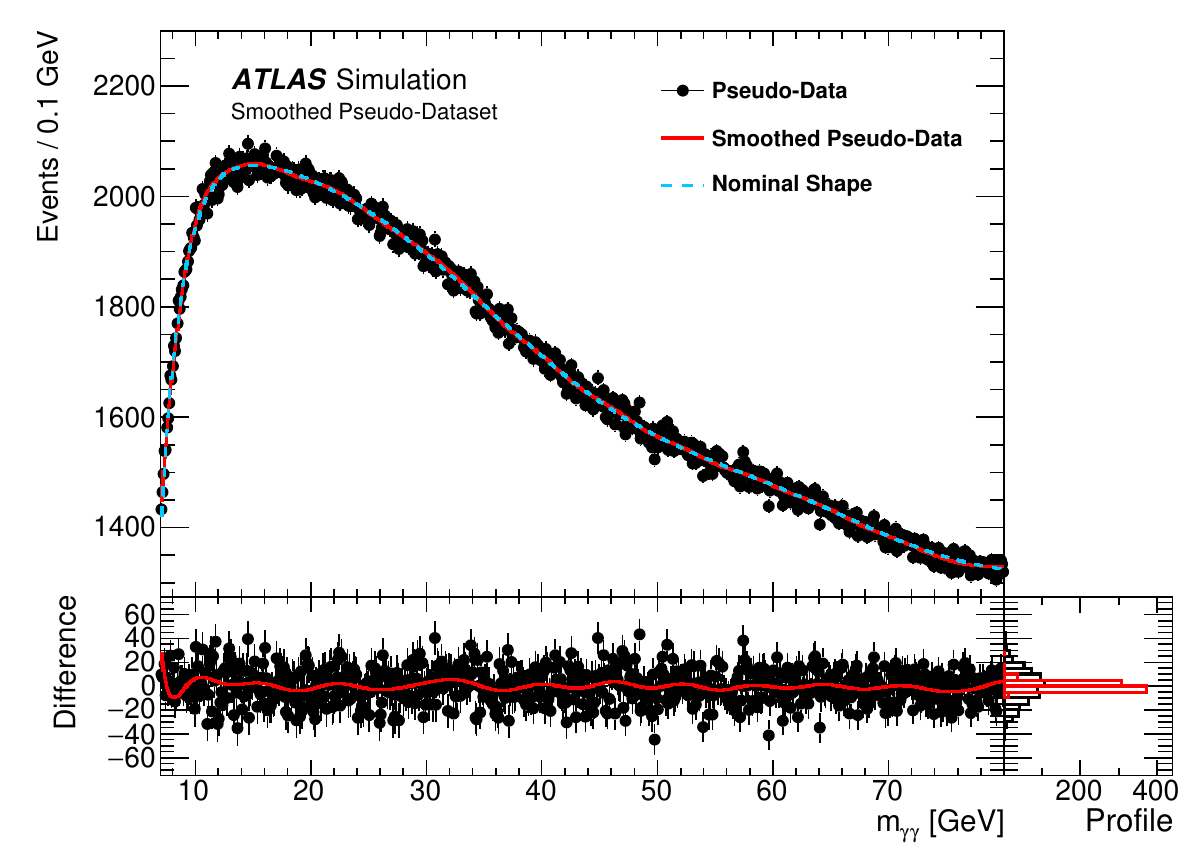}
\caption{
The upper panel shows a single pseudo-dataset (solid markers) generated from the nominal background modelling function described in Eq.~(\ref{eq:fullBkg}) (blue dashed line). The GP-smoothed pseudo-dataset is shown with the red solid line. The bottom panel shows the difference between the unsmoothed and smoothed pseudo-datasets with respect to the nominal background shape.
The horizontal axis of the plot utilizes a wider diphoton invariant mass range than the one used in the analysis in order to mitigate the impact of edge effects from the GP smoothing technique. The lower right panel shows the profile of the difference between the unsmoothed (black) and smoothed (red) pseudo-datasets with respect to the nominal background shape.
}
\label{fig:smooth_toy}
\end{figure}

\subsection{Impact of smoothing on the background modelling uncertainty}
\label{sec:sub_bkg_uncert}

In order to verify that the GP smoothing technique does not introduce any significant bias into the background histogram shape, its effect is checked on an ensemble of pseudo-datasets generated from a known background shape.
Ensembles are generated using both the nominal background modelling function (provided in Eq.~(\ref{eq:fullBkg})) and a set of analytic forms capable of describing the turn-on feature in the background shape.
This set is composed of functional forms built similarly to the nominal functional form in which either the turn-on component or the smoothly falling component is replaced by other analytic forms, such as different Fermi--Dirac-like functions for the turn-on or sums of exponential functions for the smoothly falling component.
The parameters of the analytic forms are determined by fitting them to the original simulated background sample, and each histogram in the ensemble is generated with the same effective event count as in the original simulated background sample.

The background modelling uncertainty is then evaluated for each pseudo-dataset, with and without smoothing.
The aforementioned functional forms are used to probe for potential smoothing bias in both the cases where the analytic background model did or did not properly describe the pseudo-dataset.
The bias that arises from the GP smoothing technique is defined as the difference between the observed spurious signals in the smoothed template and the unsmoothed template.
The uncertainty associated with the GP smoothing technique is observed to be roughly $20\%$ of the background modelling uncertainty for masses below 20~\GeV, stabilizing at $5\%$ for larger masses.
This bias is added in quadrature to the background modelling uncertainty.

The final background modelling uncertainty is computed as the envelope of the maximal fitted signal yields over all the background template variations defined previously in this section after smoothing.
Figure~\ref{fig:ss_plots_syst_rel} shows the number of spurious-signal events $N_\text{SS}$, taken as the background modelling uncertainty, relative to its statistical uncertainty $\delta S$ for the unsmoothed and smoothed templates.
Applying the GP smoothing procedure to the background template leads to a reduction of at least 50\% in this background modelling uncertainty relative to the unsmoothed case.
The uncertainty arising from the GP smoothing technique is found to be small compared to the decrease in background modelling uncertainty due to the reduction of statistical noise.
The magnitude of the smoothing uncertainty, as well as the remaining background modelling uncertainty, is presented as a function of the diphoton invariant mass in Figure~\ref{fig:ss_plots_syst_abs}.

\begin{figure}[h]
\centering
\subfloat[]{
\includegraphics[width=0.47\columnwidth]{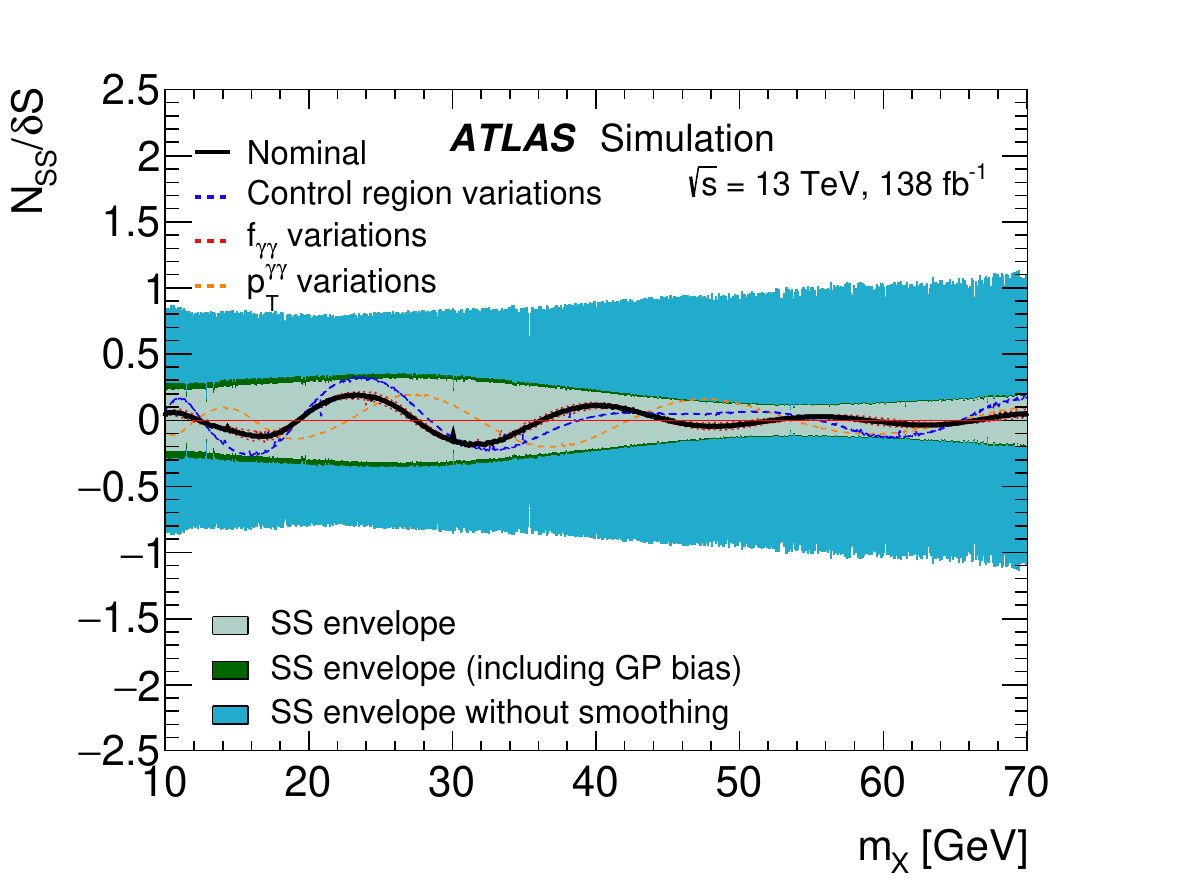}
\label{fig:ss_plots_syst_rel}
}
\subfloat[]{
\includegraphics[width=0.47\columnwidth]{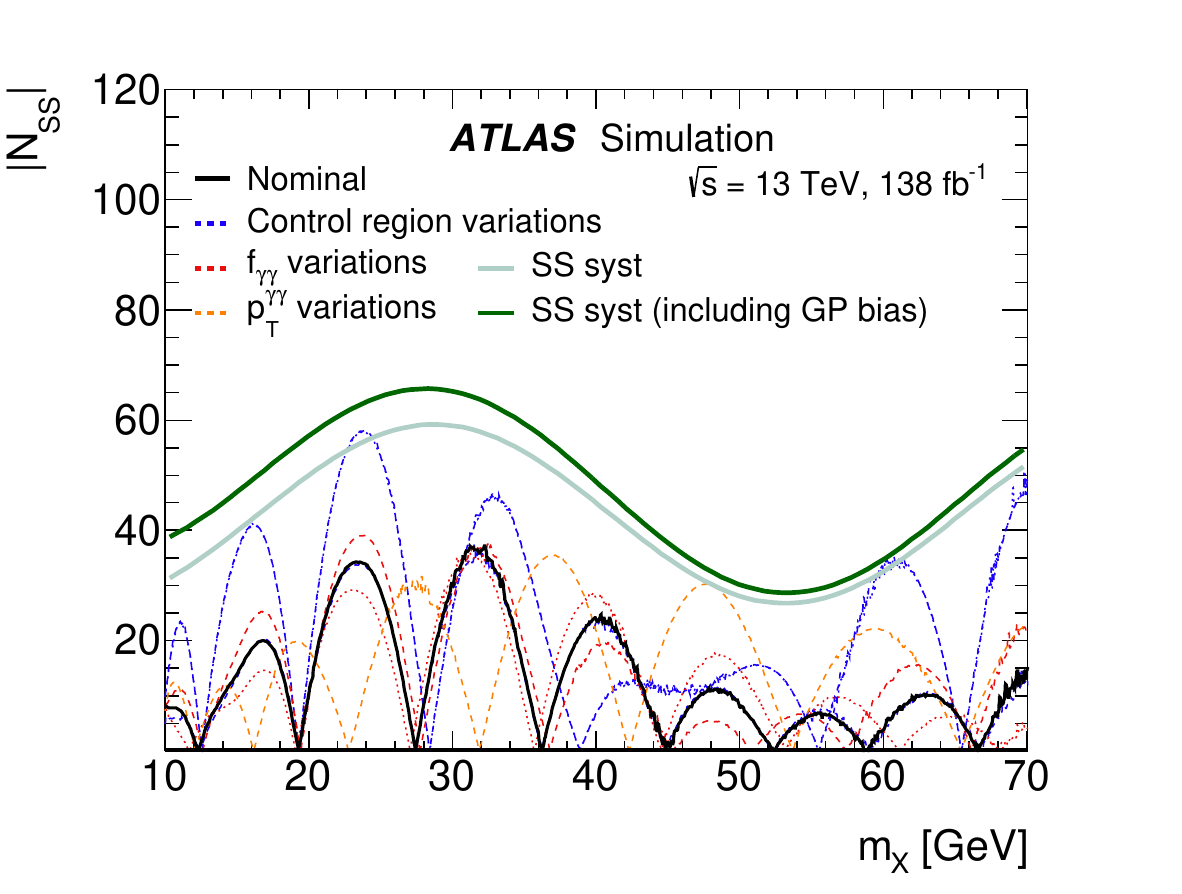}
\label{fig:ss_plots_syst_abs}
}
\hfill
\caption{
(a) Number of spurious-signal events $N_\text{SS}$ relative to its uncertainty  $\delta S$ as a function of the resonance mass $\mX$ for smoothed variations of the nominal template obtained from different control region definitions, diphoton purity variations and different kinematic selections.
The light green line encloses the local maxima of the distribution.
The dark green envelope includes the bias observed from the GP fit.
The light blue envelope shows the relative background modelling uncertainty obtained from a parameterization of the unsmoothed template.
(b) Absolute number of spurious-signal events $|N_\text{SS}|$ as a function of \mX\ evaluated for smoothed variations of the nominal template obtained from different control region definitions, diphoton purity variations and different kinematic selections.
The light green line encloses  the local maxima of the distribution.
The dark green line includes the bias observed from the GP fit.
}
\end{figure}


%
%
%

%

%
%
%
%
%
%

\section{Statistical analysis}
\label{sec:statistical}

The data are interpreted by following the statistical procedure described in Ref.~\cite{Cowan:2010js}.
A binned likelihood function is built from the observed diphoton invariant mass distribution and the analytic functions discussed in Sections~\ref{sec:signalModelling} and~\ref{sec:bkgModelling}, describing the signal and background components in the 9 to 77~\GeV\ mass range.
The search is performed in the 10 to 70~\GeV\ mass range to avoid edge effects, based on the different diphoton invariant mass resolutions at these values, as illustrated in Figure~\ref{fig:sigma_param}.

The parameter of interest to be extracted from the likelihood fit is the fiducial production cross-section times branching ratio~\sfidbr.
Since the measurement is performed in a fiducial volume (defined in Section~\ref{sec:selection}) to allow easier reinterpretation of the results, the fiducial cross-section includes a correction factor $C_{X}$ to account for the signal detection efficiency:

\begin{equation*}
\sfidbr  = \frac{N_{\text{S}}}{C_{X}\mathcal{L}}, \text{ with }\hspace{2mm} C_{X} = \frac{N^{\text{det}}_{\text{MC}}}{N^{\text{fid}}_{\text{MC}}},
\end{equation*}
where $N_{\text{S}}$ is the number of signal events fitted in data, $\mathcal{L}$ is the integrated luminosity, $N^{\text{det}}_{\text{MC}}$ is the number of reconstructed and selected signal events in the simulation and $N^{\text{fid}}_{\text{MC}}$ is the number of simulated signal events present within the fiducial volume.
The $C_{X}$ values are computed from the simulated signal samples described in Section~\ref{sec:data_mc} and range from 0.2 to 0.5 as a function of \mX.

The theoretical uncertainties affecting the measurement of \sfidbr~arise from variations of the renormalization and factorization scales affecting the signal efficiencies evaluated in simulated samples.
The experimental uncertainties directly impacting the signal yield include those involved in the luminosity determination, the modelling of pile-up interactions in simulation, the trigger efficiency, and photon identification and isolation.
An additional systematic uncertainty in the trigger is included to account for the capability of the trigger system to identify two closely spaced electromagnetic showers. Events containing a $Z\rightarrow ll\gamma$ decay, recorded only with electron triggers, in which the photon is close to one of the two electrons are used to evaluate the photon trigger efficiency in data and simulated radiative Z samples~\cite{TRIG-2018-05}. The observed difference is added in quadrature to the nominal trigger systematic uncertainty.
Uncertainties in the signal shape parameterization from the modelling and the determination of the photon energy resolution and scale are also accounted for, with mild impact on the signal yield.

The systematic uncertainties are implemented in the likelihood function as nuisance parameters constrained by Gaussian penalty terms, except for the background modelling systematic uncertainty, which is implemented as an additional signal component.
All sources of systematic uncertainties are summarized in Table~\ref{tab:systs}.

The compatibility of the observed data and the background-only hypothesis for a given signal hypothesis \mX\ is tested by estimating a local $p$-value based on a profile-likelihood-ratio test statistic, detailed in Ref.~\cite{Cowan:2010js}.
The global significance of a given event excess is computed using background-only generated pseudo-datasets to account for the look-elsewhere effect~\cite{Gross:2010qma}.

In the absence of a signal, the expected and observed 95\% confidence level (CL) exclusion limits on the cross-section times branching ratio are evaluated using the modified frequentist approach CL$_\text{s}$~\cite{Junk:1999kv,Read:2002hq} with the asymptotic approximation to the test-statistic distribution.

\begin{table}[!h]
\centering
\caption{Summary of the main sources of systematic uncertainty.
Their impact on the fiducial production cross-section of a hypothetical resonant signal is shown, except for the background modelling uncertainty, which is expressed both as a number of events and relative to the expected statistical uncertainty $\delta S$ of a fitted signal.
Unless written otherwise, numbers are mass independent.
}
\begin{tabular}{lc}
\hline \hline
Source                   & Uncertainty  \\
\hline
\multicolumn{2}{c}{On \sfidbr~[\%]}                                           \\
\hline\hline
Pile-up modelling & $\pm 3.5$ (at 10~\GeV) to $\pm 2$ (beyond 15~\GeV), mass dependent\\
Photon energy resolution & $\pm 2.5$ to $\pm 2.7$, mass dependent \\
Scale and PDFs uncertainties & $\pm 2.5$ to $\pm 0.5$, mass dependent \\
Trigger on closely spaced photons & $\pm 2$ (at 10~\GeV) to ${<}0.1$ (beyond 35~\GeV), mass dependent \\
Photon identification    & $\pm 2.0 $ \\
Isolation efficiency     & $\pm 2.0 $ \\
Luminosity (2015--2018)  & $\pm 1.7 $ \\ %
Trigger                  & $\pm 1.0  $ \\
Signal shape modelling & ${<}1$~~~\, \\
Photon energy scale & negligible \\
\hline
\multicolumn{2}{c}{\em Background modelling}                                           \\
\hline \hline
Spurious signal (relative to $\delta S$)& 30--65 events (10\%--30\%), mass dependent \\
\hline \hline
\end{tabular}
\label{tab:systs}
\end{table}


%
%
%

\section{Results}

The diphoton invariant mass distribution of events passing the analysis selection is shown in Figure~\ref{fig:inv_mass_data}, along with the background-only fit performed in the 9 to 77~\GeV\ mass range.

\begin{figure}[!h]
\centering
\subfloat[]{
\includegraphics[width=0.45\columnwidth]{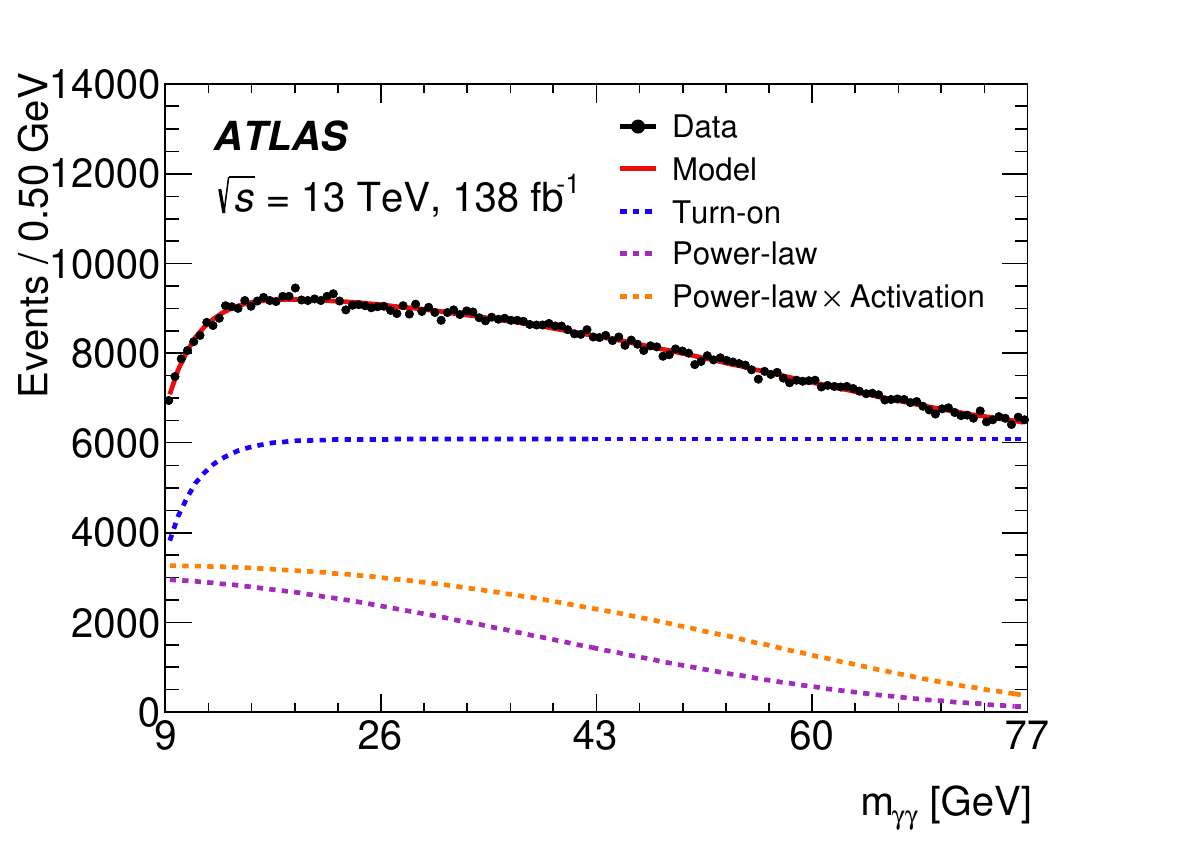}
\label{fig:inv_mass_data_decomp}
}
\subfloat[]{
\includegraphics[width=0.45\columnwidth]{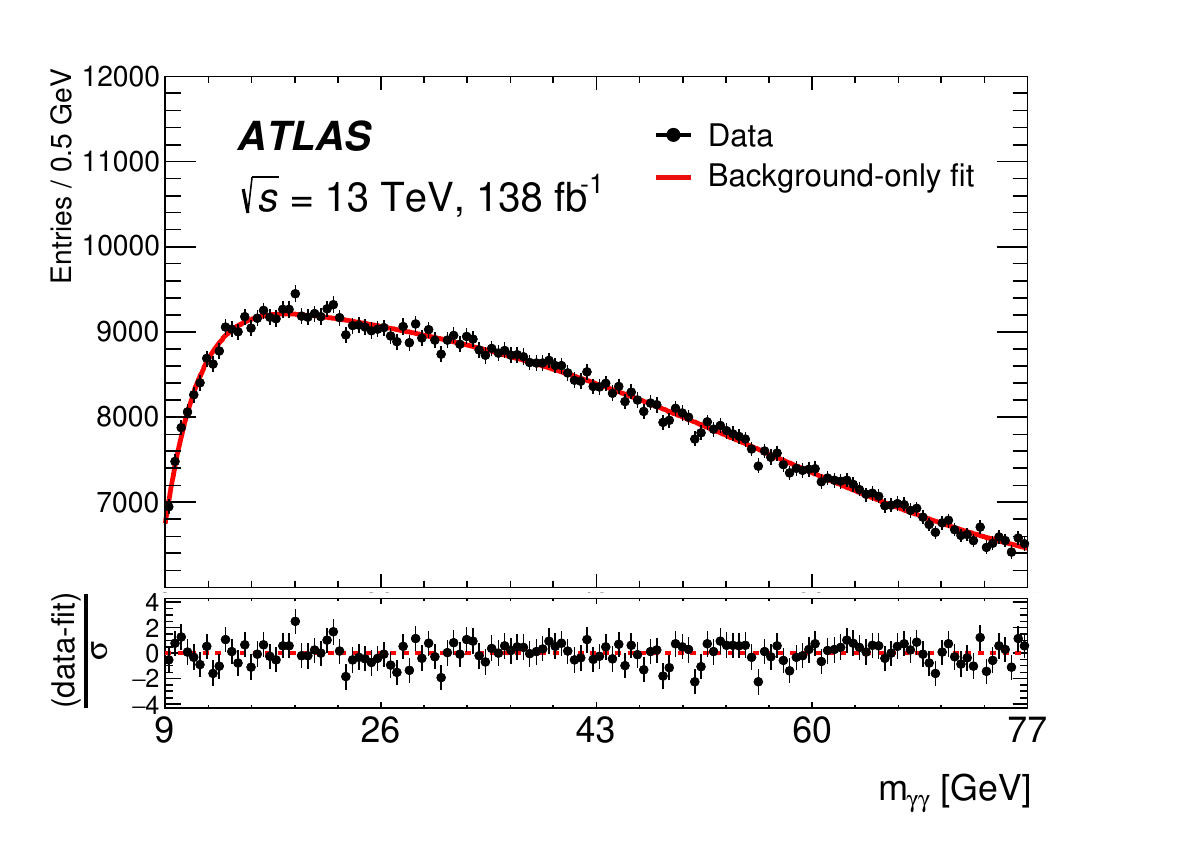}
}
\hfill
\caption{
(a)  Distribution of the diphoton invariant mass for all events passing the analysis selections in the full Run~2 dataset with the background-only fit superimposed.
The functional form is decomposed into the different pieces described in detail in Section~\ref{sec:sub_bkg_model}.
(b) Vertically enlarged version of Figure~\ref{fig:inv_mass_data_decomp}.
The normalized residuals between the data and the fit are shown in the bottom panel.
}
\label{fig:inv_mass_data}
\end{figure}

The result of the $p$-value scan as a function of the hypothesized resonance mass \mX~is shown in Figure~\ref{fig:pval_scan}.
The most significant deviation from the background-only hypothesis is observed for a mass of 19.4~\GeV, corresponding to a local significance of $3.1\sigma$.
The global significance of such an excess is $1.5\sigma$, computed using the methodology described in Section~\ref{sec:statistical}. %
Therefore, no significant deviations from the SM are observed.

\begin{figure}[!h]
\centering
\includegraphics[width=0.8\columnwidth]{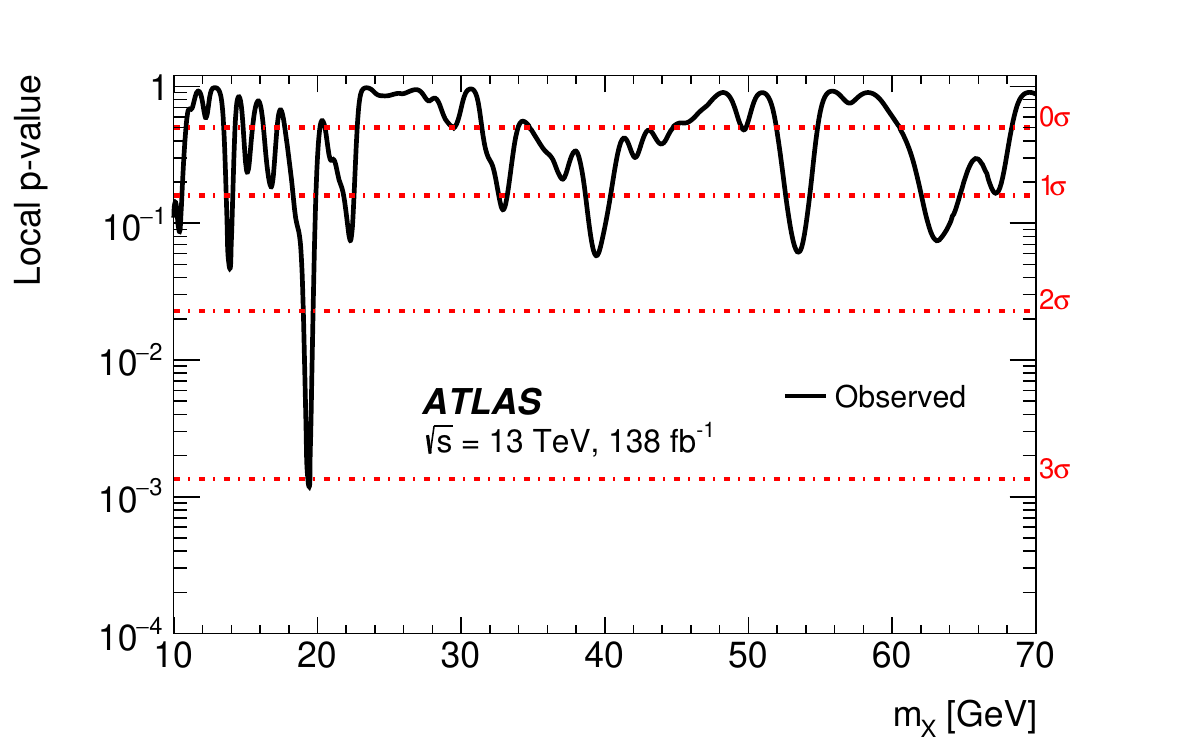}
\caption{Scan of the observed $p$-value as a function of the diphoton invariant mass for the background-only hypothesis for the full Run~2 dataset.
}
\label{fig:pval_scan}
\end{figure}

The observed and expected upper limits on \sfidbr~as a function of the resonance mass are shown in Figure~\ref{fig:scan_mass}.
The expected upper limit is nearly constant across almost the whole invariant mass range of the search because of the competing effects of the increase of signal efficiencies and widening of the signal widths with increasing mass.

\begin{figure}[!ht]
\centering
\includegraphics[width=0.8\columnwidth]{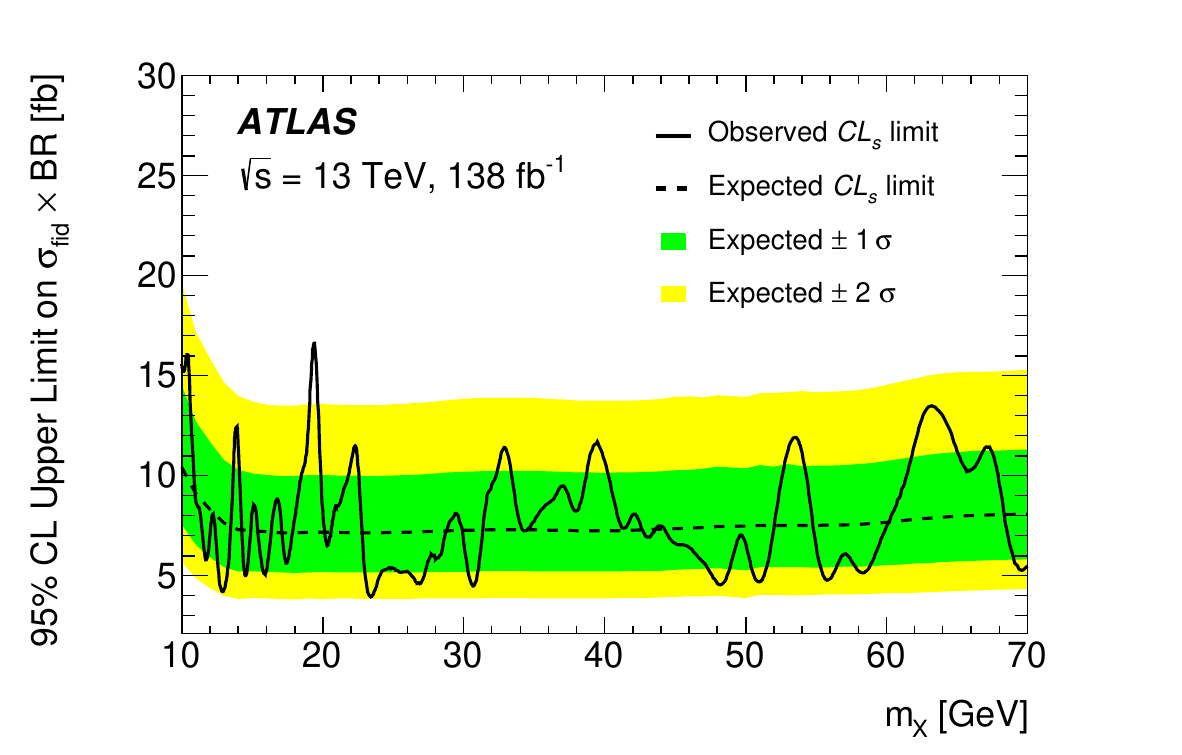}
\hfill
\caption{
Expected and observed upper limits on the fiducial production cross-section times branching ratio to photon pairs for a scalar resonance in the NWA as a function of the resonance mass \mX, for the full Run~2 dataset.
}
\label{fig:scan_mass}
\end{figure}


\section{Phenomenological interpretation}

In this section the observed limits on the fiducial cross-section for a hypothetical resonance $X$ are recast in the parameter space of an ALP $a$.
The KSVZ-ALP model, inspired by the simplest QCD-axion model~\cite{Peccei:1977hh,Kim:1979if,Shifman:1979if}, is chosen as a benchmark because it allows for couplings between the ALP field and gauge bosons, including a non-zero coupling to gluons and photons. It is described  by the following effective Lagrangian:
\begin{equation}
\frac{a}{4\pi f_a}\left[ \alpha_3 c_3 G^a\tilde{G}^a+ \alpha_2 c_2 W^i\tilde{W}^i+ \alpha_1 c_1 B\tilde{B}\right] + \frac{1}{2}m^{2}_{a}a^{2}\ ,
\label{eq:LaFFdual}
\end{equation}
where $a$ and $m_{a}$ are the ALP field and its mass, and $f_{a}$ corresponds to the decay constant that governs its coupling with the SM fields.
The QCD and EW field strengths are denoted by $G$, $W$ and $B$ with $\tilde{F}^{\mu\nu}=(1/2)\,\epsilon^{\mu\nu\rho\sigma}F_{\rho\sigma}$ for all field strengths.
The coupling constants $\alpha_{3}$ , $\alpha_{2}$ and $\alpha_1=(5/3)\, \alpha_{Y}$ (where $\alpha_{Y}$ stands for the weak SM hypercharge $Y$ coupling constant) set the strength of the strong and EW interactions in the SM.
The coefficients $c_{i}$ encode the anomalies of the global symmetry non-linearly realized by the ALP with the SM gauge group.
These anomalies are generated by integrating out heavy fermions which are charged under the SM gauge group at the scale $f_{a}$.
This Lagrangian is equivalent to the one used to generate the simulated signal samples described in Section~\ref{sec:data_mc}.

The ALP $a$ under consideration, being the pNGB of an approximate global symmetry, remains naturally light well below the scale of new physics.
Considering $m_{a}$, the mass of the ALP, to be much smaller than $m_Z$, the relevant two-body decays of $a$ are to photons and to jets, with widths which can be found, e.g., in Ref.~\cite{Mariotti:2017vtv}.
In the 10 to 70~\GeV\ mass range and for the choice of anomalies $c_{1}=c_{2}=c_{3}=10$, the branching ratio $\mathcal{B}(a\rightarrow\gamgam)$ varies from $0.6\cdot10^{-3}$ to $1.6\cdot10^{-3}$.
Choosing to set the magnitude of the $c_i$ parameters to be the same is motivated by gauge coupling unification in a Grand Unified Theory scenario.
While the specific value of 10 is arbitrary, the rescaling of the results to a different anomaly parameter choice would be trivial.
The ALP Lagrangian in Eq.~\eqref{eq:LaFFdual} is implemented in Feynrules~\cite{Alloul:2013bka}, and the production cross-section at the LHC for the process $gg\rightarrow ag$ is computed at leading order with \MADGRAPH~\cite{Alwall:2011uj}, where the gluon is explicitly required to boost the ALP.
A constant $K$-factor $K=2$ is applied to this cross-section to account for NLO corrections, which were computed for a similar signal topology in Ref.~\cite{Gershtein:2020mwi}.
ALPs that couple to gluons decay promptly over the entire mass range of interest for this study (recent studies of displaced ALP decays can be found in Refs.~\cite{Hook:2019qoh,Gershtein:2020mwi}).
Because of the large hierarchy between $f_a$ and $m_a$, and the loop suppression of the coupling, the ALP total width is dominated by its coupling to gluons and is always small compared to its mass.
As a consequence, the narrow-width approximation always applies and finite-width effects can be safely neglected.

\begin{figure}[t]
\centering
\includegraphics[scale=0.7]{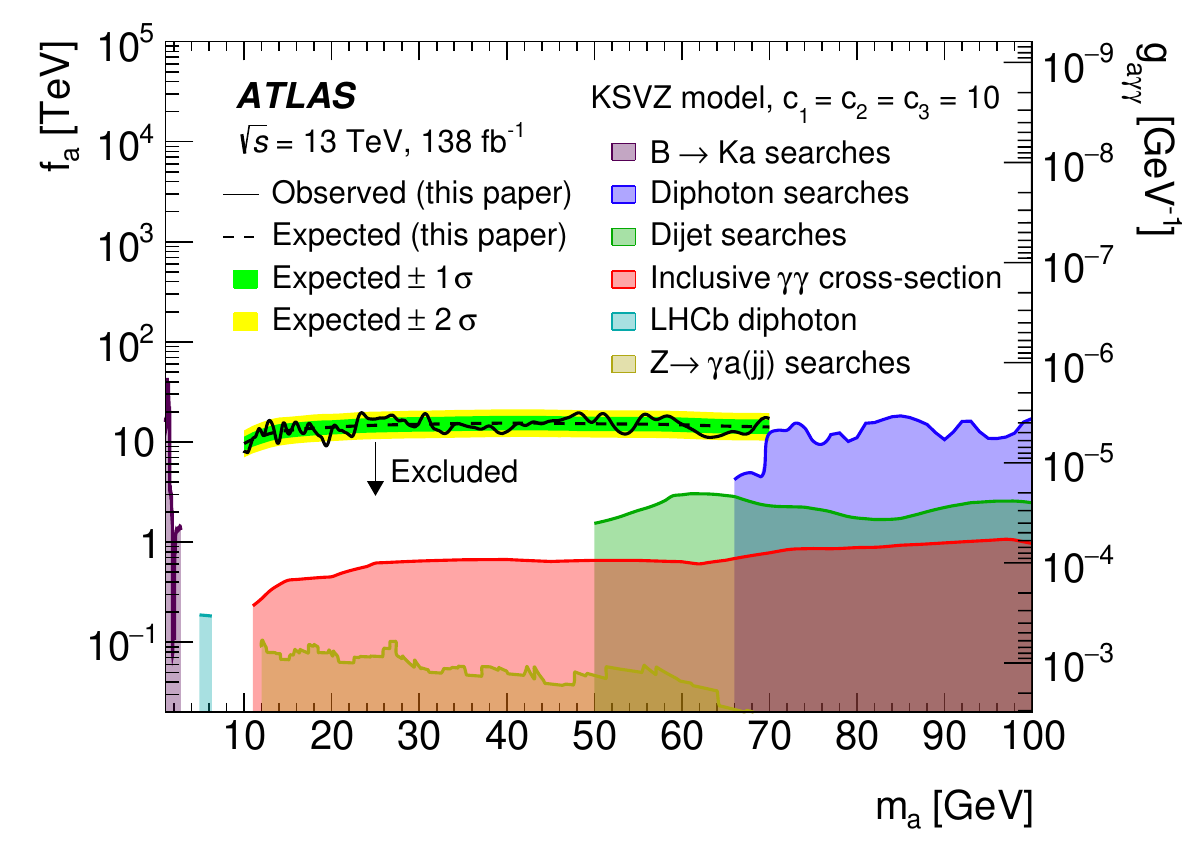}
\caption{
Parameter space of the ALP for $c_1=c_2=c_3=10$ (Eq.~(\ref{eq:LaFFdual})).
The observed and expected lower bounds on the ALP decay constant derived from this analysis are shown in black solid and dashed lines respectively.
\BABAR\ bounds on $B\to K a$ derived in Ref.~\cite{Chakraborty:2021wda} are  shown in purple; in green the LHC bounds on boosted dijet resonances~\cite{CMS-EXO-17-001} and in blue the LHC searches for diphoton resonances taken from Ref.~\cite{Mariotti:2017vtv}.
The red bounds are derived from Tevatron~\cite{Aaltonen:2012jd} and LHC~\cite{CMS-SMP-13-001,STDM-2012-05,STDM-2017-30} diphoton cross-section measurements, following the method described in Ref.~\cite{Mariotti:2017vtv}.
Weaker constraints covering lower invariant masses are obtained from LHCb diphoton measurements~\cite{CidVidal:2018blh} and from LEP searches for $Z\to \gamma a(jj)$~\cite{Adriani:1992zm}, in cyan and yellow respectively.
On the right, the $y$-axis shows the ALP--photon coupling %
$g_{a\gamma\gamma}\equiv \alpha_{\text{em} }E/\pi f_{a}$ ($E=c_2+\frac{5}{3}c_1$), a standard QCD axion notation.}
\label{fig:ma_vs_fa}
\end{figure}

The recasting is done by comparing the theoretical signal yield obtained from the ALP model of Eq.~(\ref{eq:LaFFdual}), after applying the particle-level selection described in Section~\ref{sec:selection}, with the bounds on the fiducial cross-section in Figure~\ref{fig:scan_mass}.
The signal cross-section times branching ratio can be written as $1/f^{2}_{a}$ times a weakly varying function of the ALP mass. The upper limit on the cross-section then results in a lower limit on $f_{a}$, which is shown in Figure~\ref{fig:ma_vs_fa} for a specific choice of the $c_i$ coefficients.

Figure~\ref{fig:ma_vs_fa} shows how the sensitivity of the search presented here covers a large portion of the unexplored ALP parameter space where the heavy colour states generating the ALP coupling to gauge bosons are in the multi-\TeV\ range and therefore unaccessible at the LHC.
Any production mechanism other than gluon--gluon fusion suffers from a smaller production cross-section, and the decoupling of the heavy states inducing the ALP coupling to SM states would require further study.

Constraints from $\Upsilon\to \gamma a(jj)$~\cite{Lees:2011wb}, constraints from  $Z$ boson width measurements~\cite{Workman:2022ynf}, and ALP production in light-by-light scattering in heavy-ion collisions~\cite{HION-2019-08,CMS-FSQ-16-012} are too weak to appear in the plot.


%

%
%

%

\section{Conclusion}

A search for new narrow-width boosted resonances is performed in the diphoton invariant mass spectrum ranging from 10 to 70~\GeV, using 138~\ifb\ of $pp$ collision data collected at a centre-of-mass energy of 13~\TeV\ with the ATLAS detector at the Large Hadron Collider.
The data are consistent with the SM background expectation.
Limits are set on the fiducial cross-section times branching ratio in a fiducial region defined to mimic the detector-level selection.
The observed limits on \sfidbr~range from 4 to 17~fb, with variations mainly due to statistical fluctuations of the data.
The dominant uncertainties arise from the limited number of $pp$ collisions collected and the background modelling uncertainty.
The impact of the latter is reduced by smoothing the simulated background-only sample with Gaussian Processes in order to reduce the statistical fluctuations in the sample.
Furthermore, the observed limits are recast in the parameter space of an axion-like particle, covering a longstanding gap in diphoton resonance searches.


%

\FloatBarrier

\section*{Acknowledgements}
%

%
%

%
%

We thank CERN for the very successful operation of the LHC, as well as the
support staff from our institutions without whom ATLAS could not be
operated efficiently.

We acknowledge the support of
ANPCyT, Argentina;
YerPhI, Armenia;
ARC, Australia;
BMWFW and FWF, Austria;
ANAS, Azerbaijan;
CNPq and FAPESP, Brazil;
NSERC, NRC and CFI, Canada;
CERN;
ANID, Chile;
CAS, MOST and NSFC, China;
Minciencias, Colombia;
MEYS CR, Czech Republic;
DNRF and DNSRC, Denmark;
IN2P3-CNRS and CEA-DRF/IRFU, France;
SRNSFG, Georgia;
BMBF, HGF and MPG, Germany;
GSRI, Greece;
RGC and Hong Kong SAR, China;
ISF and Benoziyo Center, Israel;
INFN, Italy;
MEXT and JSPS, Japan;
CNRST, Morocco;
NWO, Netherlands;
RCN, Norway;
MEiN, Poland;
FCT, Portugal;
MNE/IFA, Romania;
MESTD, Serbia;
MSSR, Slovakia;
ARRS and MIZ\v{S}, Slovenia;
DSI/NRF, South Africa;
MICINN, Spain;
SRC and Wallenberg Foundation, Sweden;
SERI, SNSF and Cantons of Bern and Geneva, Switzerland;
MOST, Taiwan;
TENMAK, T\"urkiye;
STFC, United Kingdom;
DOE and NSF, United States of America.
In addition, individual groups and members have received support from
BCKDF, CANARIE, Compute Canada and CRC, Canada;
PRIMUS 21/SCI/017 and UNCE SCI/013, Czech Republic;
COST, ERC, ERDF, Horizon 2020 and Marie Sk{\l}odowska-Curie Actions, European Union;
Investissements d'Avenir Labex, Investissements d'Avenir Idex and ANR, France;
DFG and AvH Foundation, Germany;
Herakleitos, Thales and Aristeia programmes co-financed by EU-ESF and the Greek NSRF, Greece;
BSF-NSF and MINERVA, Israel;
Norwegian Financial Mechanism 2014-2021, Norway;
NCN and NAWA, Poland;
La Caixa Banking Foundation, CERCA Programme Generalitat de Catalunya and PROMETEO and GenT Programmes Generalitat Valenciana, Spain;
G\"{o}ran Gustafssons Stiftelse, Sweden;
The Royal Society and Leverhulme Trust, United Kingdom.

The crucial computing support from all WLCG partners is acknowledged gratefully, in particular from CERN, the ATLAS Tier-1 facilities at TRIUMF (Canada), NDGF (Denmark, Norway, Sweden), CC-IN2P3 (France), KIT/GridKA (Germany), INFN-CNAF (Italy), NL-T1 (Netherlands), PIC (Spain), ASGC (Taiwan), RAL (UK) and BNL (USA), the Tier-2 facilities worldwide and large non-WLCG resource providers. Major contributors of computing resources are listed in Ref.~\cite{ATL-SOFT-PUB-2023-001}.

%
%


%
%
%
%
%
%

%
\clearpage
\appendix
\printbibliography
\clearpage

 
\begin{flushleft}
\hypersetup{urlcolor=black}
{\Large The ATLAS Collaboration}

\bigskip

\AtlasOrcid[0000-0002-6665-4934]{G.~Aad}$^\textrm{\scriptsize 101}$,
\AtlasOrcid[0000-0002-5888-2734]{B.~Abbott}$^\textrm{\scriptsize 119}$,
\AtlasOrcid[0000-0002-7248-3203]{D.C.~Abbott}$^\textrm{\scriptsize 102}$,
\AtlasOrcid[0000-0002-1002-1652]{K.~Abeling}$^\textrm{\scriptsize 55}$,
\AtlasOrcid[0000-0002-8496-9294]{S.H.~Abidi}$^\textrm{\scriptsize 29}$,
\AtlasOrcid[0000-0002-9987-2292]{A.~Aboulhorma}$^\textrm{\scriptsize 35e}$,
\AtlasOrcid[0000-0001-5329-6640]{H.~Abramowicz}$^\textrm{\scriptsize 150}$,
\AtlasOrcid[0000-0002-1599-2896]{H.~Abreu}$^\textrm{\scriptsize 149}$,
\AtlasOrcid[0000-0003-0403-3697]{Y.~Abulaiti}$^\textrm{\scriptsize 116}$,
\AtlasOrcid[0000-0003-0762-7204]{A.C.~Abusleme~Hoffman}$^\textrm{\scriptsize 136a}$
% ...
\bigskip
\\

% ...
$^{29}$Physics Department, Brookhaven National Laboratory, Upton NY; United States of America.\\
% ...
$^{35}$$^{(a)}$Facult\'e des Sciences Ain Chock, R\'eseau Universitaire de Physique des Hautes Energies - Universit\'e Hassan II, Casablanca;$^{(b)}$Facult\'{e} des Sciences, Universit\'{e} Ibn-Tofail, K\'{e}nitra;$^{(c)}$Facult\'e des Sciences Semlalia, Universit\'e Cadi Ayyad, LPHEA-Marrakech;$^{(d)}$LPMR, Facult\'e des Sciences, Universit\'e Mohamed Premier, Oujda;$^{(e)}$Facult\'e des sciences, Universit\'e Mohammed V, Rabat;$^{(f)}$Institute of Applied Physics, Mohammed VI Polytechnic University, Ben Guerir; Morocco.\\
% ...
$^{55}$II. Physikalisches Institut, Georg-August-Universit\"{a}t G\"ottingen, G\"ottingen; Germany.\\
% ...
$^{101}$CPPM, Aix-Marseille Universit\'e, CNRS/IN2P3, Marseille; France.\\
$^{102}$Department of Physics, University of Massachusetts, Amherst MA; United States of America.\\
% ...
$^{116}$Department of Physics, New York University, New York NY; United States of America.\\
% ...
$^{119}$Homer L. Dodge Department of Physics and Astronomy, University of Oklahoma, Norman OK; United States of America.\\
% ...
$^{136}$$^{(a)}$Departamento de F\'isica, Pontificia Universidad Cat\'olica de Chile, Santiago;$^{(b)}$Millennium Institute for Subatomic physics at high energy frontier (SAPHIR), Santiago;$^{(c)}$Instituto de Investigaci\'on Multidisciplinario en Ciencia y Tecnolog\'ia, y Departamento de F\'isica, Universidad de La Serena;$^{(d)}$Universidad Andres Bello, Department of Physics, Santiago;$^{(e)}$Instituto de Alta Investigaci\'on, Universidad de Tarapac\'a, Arica;$^{(f)}$Departamento de F\'isica, Universidad T\'ecnica Federico Santa Mar\'ia, Valpara\'iso; Chile.\\
% ...
$^{149}$Department of Physics, Technion, Israel Institute of Technology, Haifa; Israel.\\
$^{150}$Raymond and Beverly Sackler School of Physics and Astronomy, Tel Aviv University, Tel Aviv; Israel.\\
% ...
$^{a}$ Also Affiliated with an institute covered by a cooperation agreement with CERN.\\
$^{b}$ Also at An-Najah National University, Nablus; Palestine.\\
$^{c}$ Also at Borough of Manhattan Community College, City University of New York, New York NY; United States of America.\\
$^{d}$ Also at Bruno Kessler Foundation, Trento; Italy.\\
$^{e}$ Also at Center for High Energy Physics, Peking University; China.\\
$^{f}$ Also at Centro Studi e Ricerche Enrico Fermi; Italy.\\
% ...

\end{flushleft}


\end{document}